\documentclass[aps,prd,preprint,nofootinbib]{revtex4-1}

\usepackage{amsfonts}
\usepackage{mathrsfs}
\usepackage{graphicx}
\usepackage{amsmath}
\usepackage{amssymb}
\usepackage{epsfig}
\usepackage{slashed}
\usepackage{color}
\usepackage{multirow}
\usepackage{ulem}
\usepackage{array}

\usepackage{threeparttable}
\usepackage{booktabs}
\usepackage{adjustbox}
\usepackage{float}
\usepackage{subfig}
\usepackage{braket}
\usepackage{bm}
\usepackage{BOONDOX-cal}
\usepackage[mathscr]{eucal}
\usepackage{array}
\usepackage[figuresleft]{rotating}
\usepackage[bottom]{footmisc}
\usepackage[colorlinks,linkcolor=blue, anchorcolor=green,citecolor=blue]{hyperref}
\usepackage[capitalize]{cleveref}
\usepackage[justification=raggedright]{caption}

\begin{document}

\title{
Categorizing
 \texorpdfstring{$SU(3)_f$}{} representations
of scalar mesons by 
 \texorpdfstring{$J/\psi$}{} decays}

\author{Chao-Qiang Geng$^{1}$, Chia-Wei Liu$^{2,3}$, Xiao Yu$^{1,4,5}$\footnote[1]{yuxiao21@mails.ucas.ac.cn}, Ao-Wen Zhou$^{1,4,5}$\\}

\affiliation{$^1$School of Fundamental Physics and Mathematical Sciences, Hangzhou Institute for Advanced Study, UCAS, Hangzhou 310024, China}
\affiliation{$^2$Tsung-Dao Lee Institute, Shanghai Jiao Tong University, 
Shanghai 200240, China}
\affiliation{$^3$Shanghai Key Laboratory for Particle Physics and Cosmology,
Shanghai Jiao Tong University, Shanghai 200240, China}
\affiliation{$^4$University of Chinese Academy of Sciences,  Beijing 100190, China}
\affiliation{$^5$Institute of Theoretical Physics, Chinese Academy of Sciences, Beijing 100190, China}

\begin{abstract}
  The scalar mesons are established for a long time, but their nature is still an open question. 
  In this paper, we investigate the potential of categorizing their $SU(3)_f$ representations via $J/\psi\to SV$ and $\gamma S$, offering a criterion that may illuminate this issue. Here, $S$ ($V$) denotes scalar (vector) mesons.
  Using the $SU(3)_f$ symmetry with the current data,
  we find that $f_0(500)$ and $f_0(980)$ are mostly made of singlet and octet $SU(3)_f$ representations, respectively, 
  with the singlet-octet mixing angle of $\theta = (82.9\pm4.4)^{\circ}$. This conclusion  
 is consistent with the caculations of the quark-antiquark ($q\bar{q}$) hypothesis.
  For the scalar mesons in the range of 1-2 GeV, we discuss the mixings between $q\bar{q}$ and glueballs.
  Our  numerical results
  suggest that $f_0(1710)$ is likely composed of the scalar glueball.
We urge our experimental colleagues to measure $J/\psi \to \rho a_0(980,\ 1450,\ 1710), $ $ K^*(892)^{\pm} K^*(700,\ 1430,\ 1950)^{\mp}$ and $\omega f_0(500)$, which  provide useful information in the  $SU(3)_f$ analysis.  
   
\end{abstract}

\maketitle

\section{Introduction}

Numerous attempts have been made to comprehend the nature of scalar mesons (S)~\cite{Jaffe:1976ig,Morgan:1974cm,Montanet:1982zi,Novikov:1979va}. However, a satisfactory solution remains elusive. In the case of the lightest scalar mesons,  the existence of a nonet $SU(3)_f$ representation at mass below 1 GeV has been intensively studied~\cite{Oller:2003vf,Anisovich:1997qp,Bediaga:2003zh,Aliev:2007uu,Klempt:2021nuf,Close:2002zu,Maiani:2004uc,Amsler:2004ps,Achasov:2005hm,tHooft:2008rus,Fariborz:2009cq}. Nevertheless, its inverted mass spectrum deviates from a simple quark-antiquark ($q\bar{q}$) model, which is proven to be successful in pseudoscalar and vector mesons. This discrepancy poses a challenge to the conventional quark model. Consequently, two primary interpretations have emerged for this nonet: conventional $q\bar{q}$ states~\cite{Montanet:1982zi,Aliev:2007uu,Anisovich:1997qp,Oller:2003vf,Bediaga:2003zh} and tetraquark structures of $qq\bar{q}\bar{q}$~\cite{Jaffe:1976ig,Close:2002zu,Maiani:2004uc,Amsler:2004ps,Achasov:2005hm,tHooft:2008rus,Fariborz:2009cq,Pelaez:2003dy,Weinberg:2013cfa,Agaev:2018fvz}.

Above 1 GeV, it is  commonly believed that there is another nonet,  regarded as $q\bar{q}$ states. However, a controversy arises regarding whether these mesons represent the low-lying P-wave state of $q\bar{q}$ or the first radial excited states relative to the low-lying P-wave state~\cite{Klempt:2021nuf}. The situation gets more complex due to the following reasons: (\romannumeral1) the large number of isoscalar mesons $f_0$, (\romannumeral2) the existence of glueballs permitted by QCD~\cite{Fritzsch:1973pi,Fritzsch:1975tx}, with the lattice QCD (LQCD) predicting scalar glueball masses in the range of 1 to 1.7 GeV~\cite{Sexton:1995kd,Bali:1993fb,Morningstar:1997ff,Chen:2005mg,Athenodorou:2020ani}, and (\romannumeral3) the mixings between glueballs and $q\bar{q}$, known as hybrids.
Generally, 
$f_0(1370,\ 1500,\ 1710)$ are considered as the components of the nonet, corresponding to physical states that are mixtures of $n\bar{n}$, $s\bar{s}$ and glueballs with $n\overline{n} = (u\overline{u} +  d\overline{d})/\sqrt{2}$. However, a consensus on the mixings has not been reached.
In Refs.~\cite{Close:2005vf,Giacosa:2005zt}, the authors proposed one type of the mixing to explain the experimental data of $f_0^i$ decaying to $\pi\pi$, $K\bar{K}$, $\eta\eta$ and $\eta\eta^{\prime}$, which favors that $f_0(1370)$ is predominantly $n\bar{n}$, $f_0(1500)$ primarily a glueball mixed with $n\bar{n}$ and $s\bar{s}$, and $f_0(1710)$ dominated by $s\bar{s}$. Additionally, 
based on the quenched LQCD calculations~\cite{Lee:1999kv}, $f_0(1710)$ is suggested to be  mainly composed of the scalar glueball. This result is also hinted by those in Refs.~\cite{Frere:2015xxa,Cheng:2015iaa}.

Theoretical expectations anticipate the discovery of more isotriplet and isodoublet scalar mesons above 1~GeV~\cite{Dai:2021owu,Geng:2008gx,Wang:2022pin}.
Recently, the BaBar experiment reported the observation of   $a_0(1710)$~\cite{BaBar:2021fkz}, subsequently confirmed by the BESIII experiment~\cite{BESIII:2022npc}. 
 The presence of distinct isotriples ($a_0$) and strange isodoublets ($K^*$) beyond 1 GeV indicates the potential existence of additional $SU(3)_f$ nonets.
However, a clear picture of the structures of the nonets has not been achieved.

\begin{figure}[!htb]
    \vspace{-0.6cm} 
    \centering
    \begin{minipage}[t]{1.0\textwidth}
    \subfloat[]{\includegraphics[width=.45\textwidth]{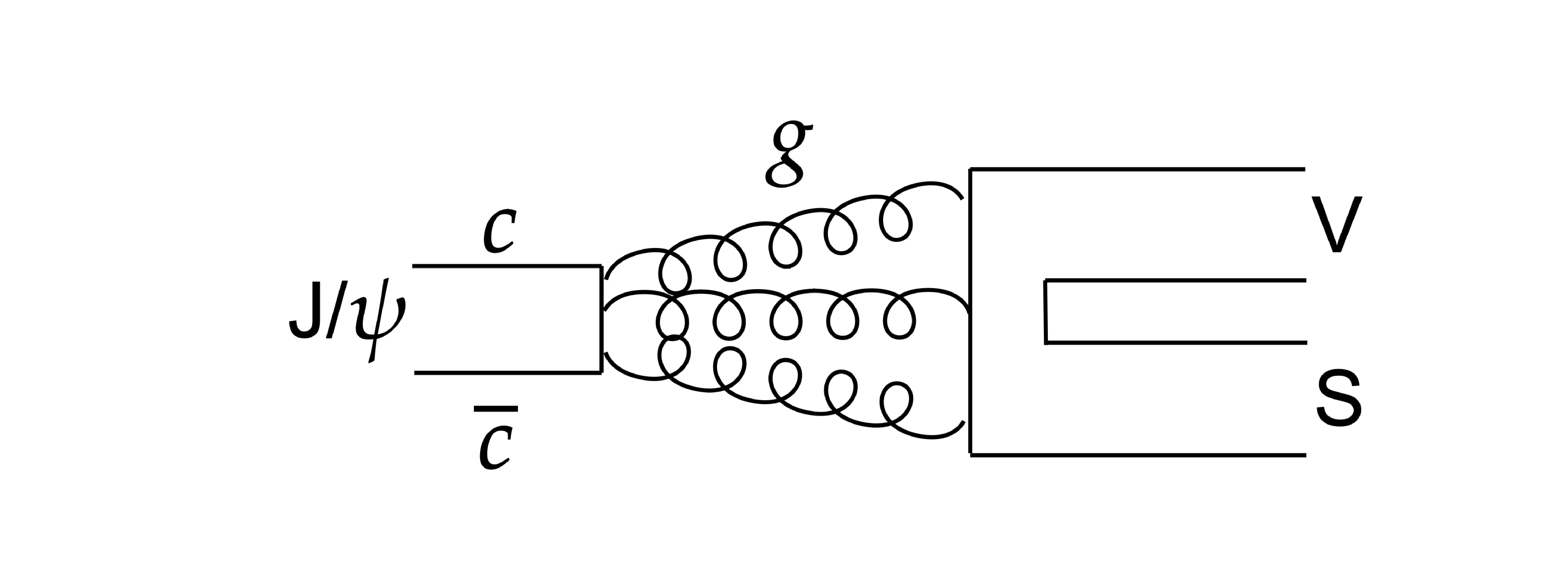}\label{fd1-a}}\quad
    \subfloat[]{\includegraphics[width=.45\textwidth]{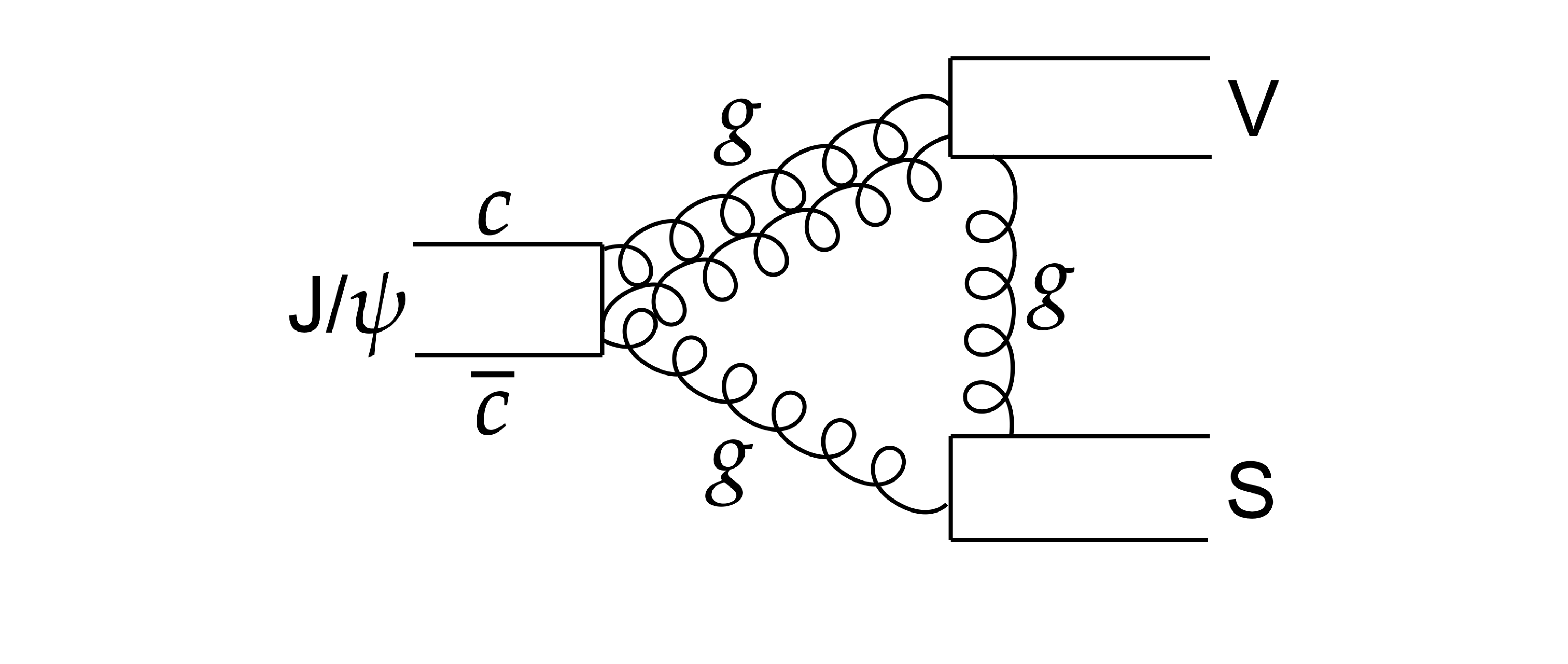}\label{fd1-b}}\\
    \subfloat[]{\includegraphics[width=.45\textwidth]{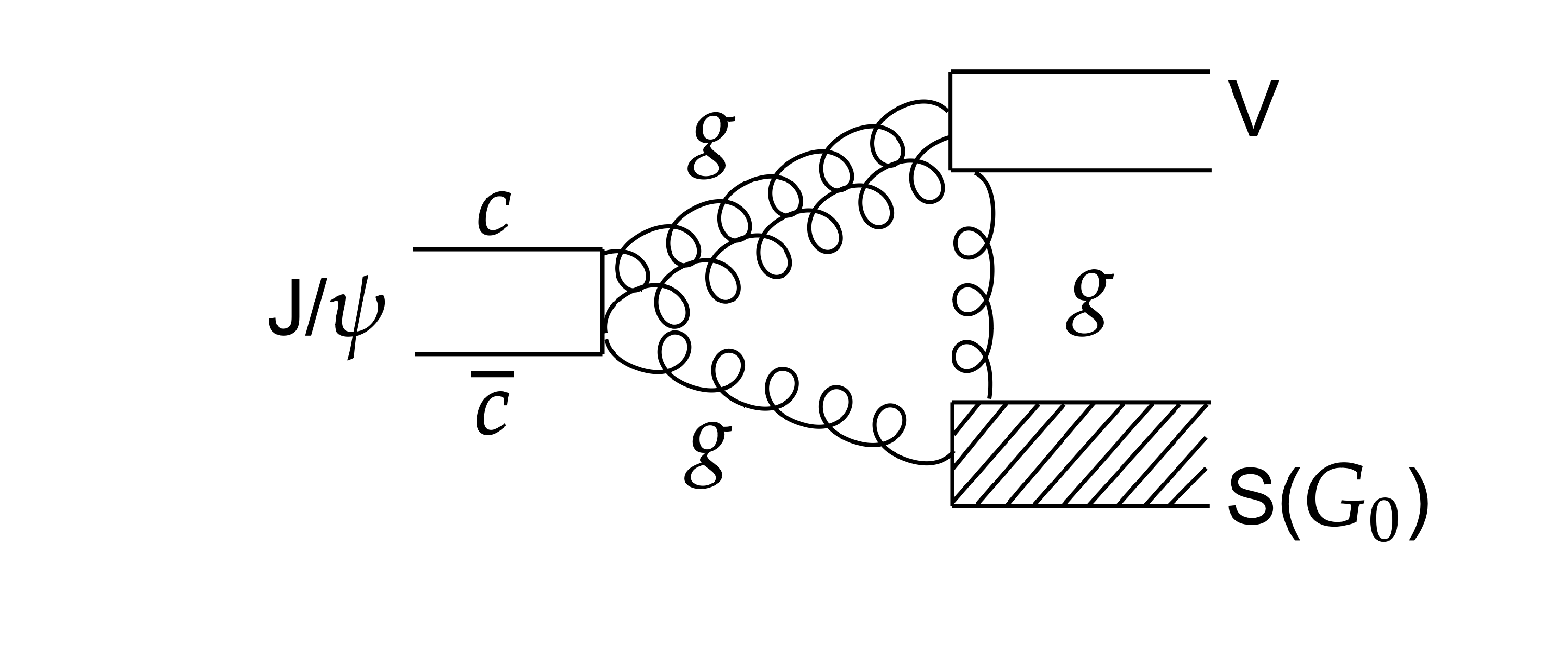}\label{fd1-c}}\quad
    \subfloat[]{\includegraphics[width=.45\textwidth]{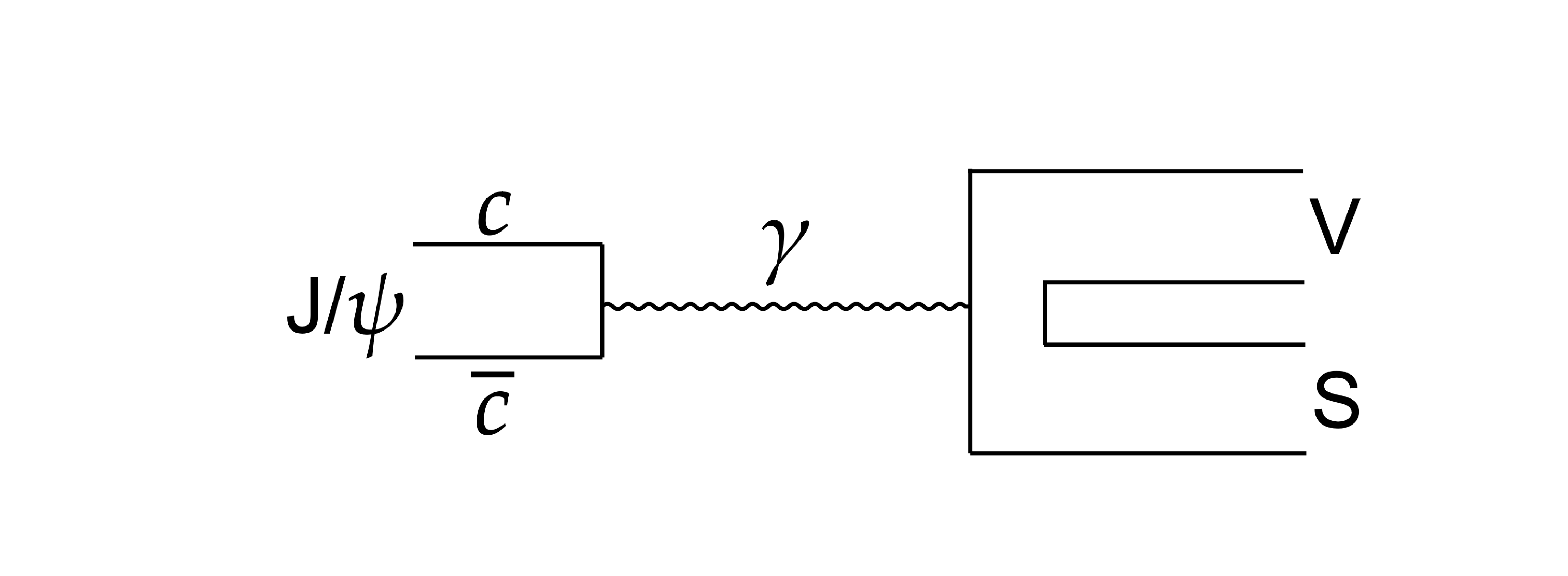}\label{fd1-d}} 
    \caption{\label{fd1} Feynman diagrams of $J/\psi\to VS$ decays with (a) SOZI-suppressed, (b) DOZI-suppressed connected to mesons, (c) DOZI-suppressed connected to a pure glueball state, and (d) EM.}
    \end{minipage}
    
    \begin{minipage}[t]{1.0\textwidth}
    \centering
    \subfloat[]{\includegraphics[width=.45\textwidth]{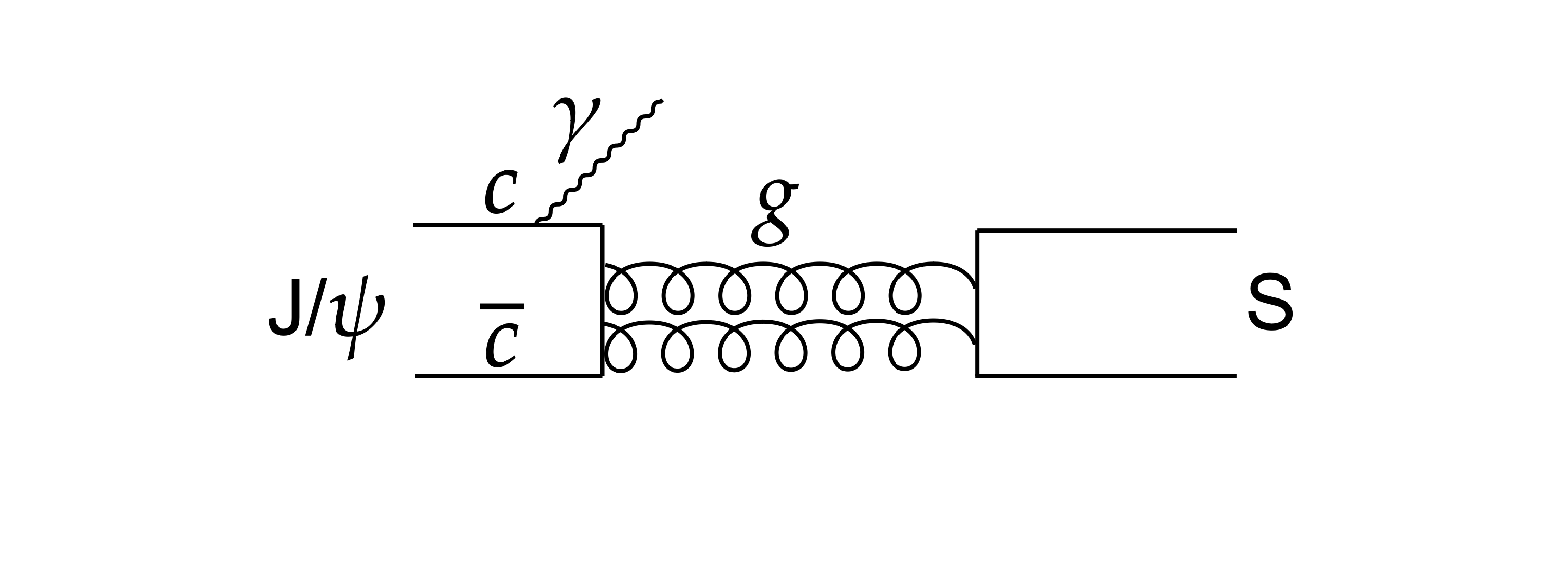}\label{fd2-a}}\quad
    \subfloat[]{\includegraphics[width=.45\textwidth]{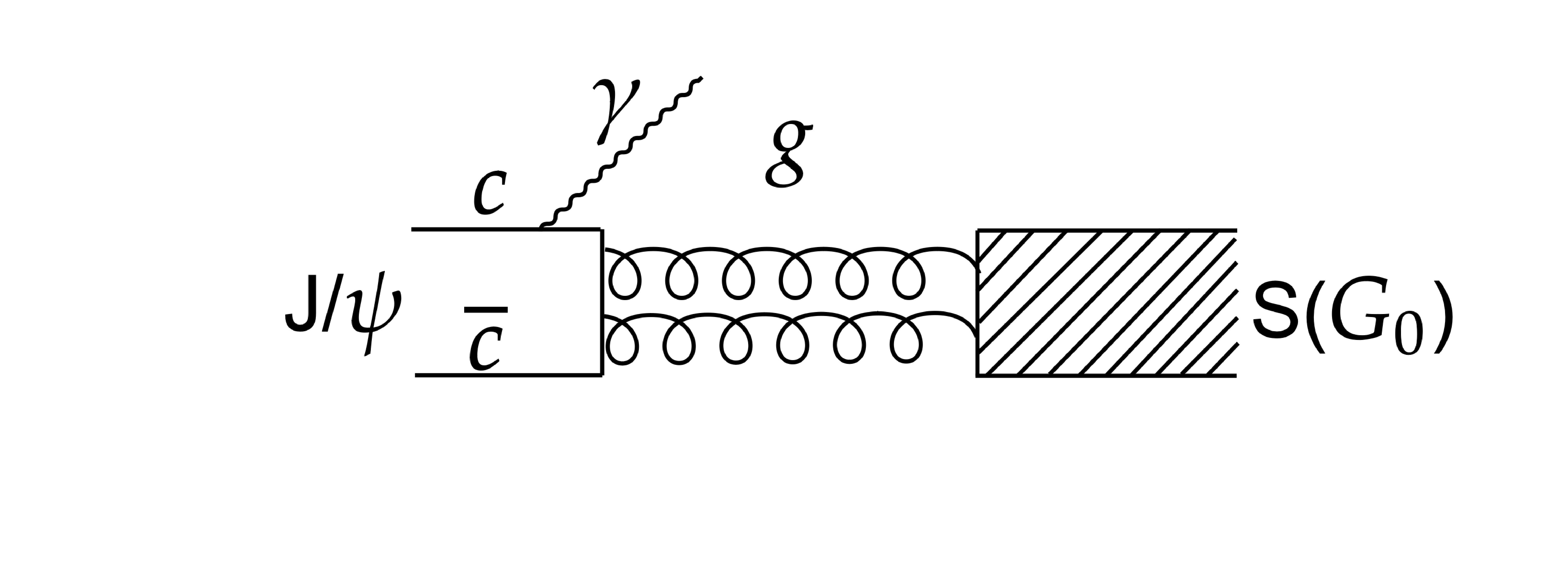}\label{fd2-b}}\\
    \subfloat[]{\includegraphics[width=.45\textwidth]{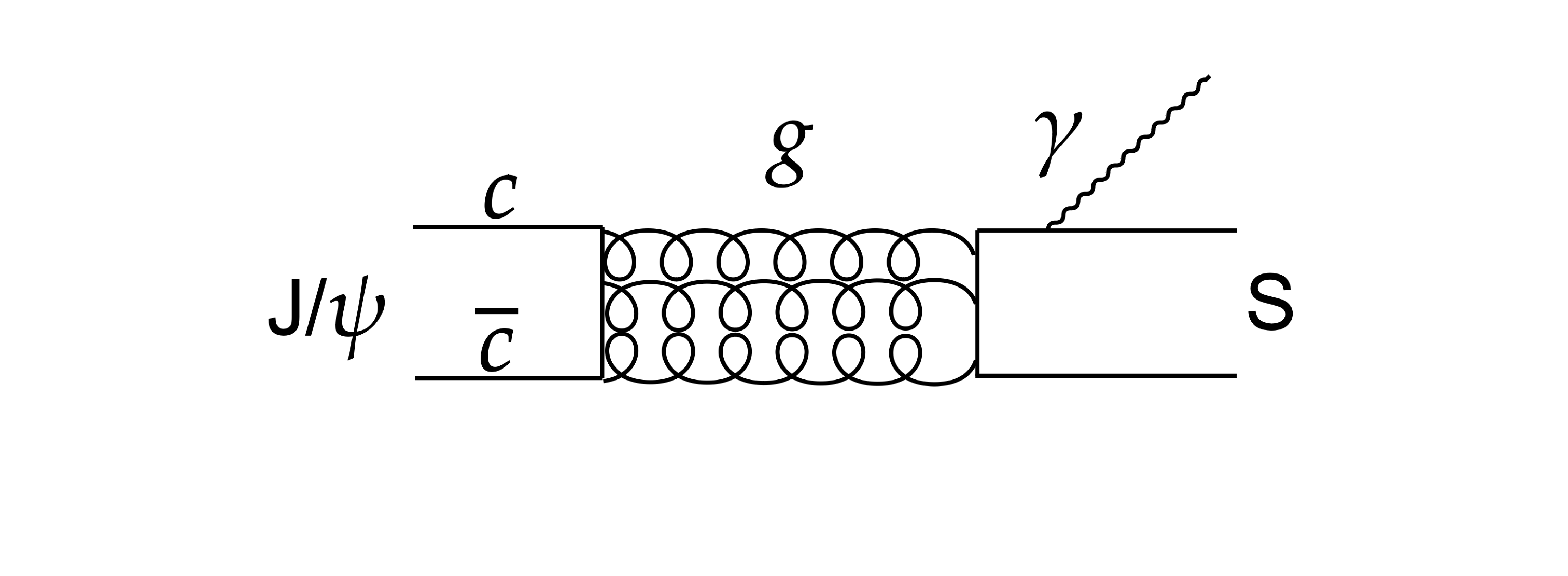}\label{fd2-c}}
    \caption{\label{fd2} Feynman diagrams of $J/\Psi\to\gamma S$ decays with (a) two-gluon-annihilation connected to mesons, (b) two-gluon-annihilation connected to a pure glueball state, and (c) three-gluon-annihilation.}
    \end{minipage}
\end{figure}

The study of $J/\psi$ decays is an useful tool for examining hadronic nonets. Since $J/\psi$ is an $SU(3)_f$ singlet state, the final states should exhibit $SU(3)_f$ invariant characteristics. 
Some attempts have been proposed to understand $J/\psi \to P V$ processes~\cite{Kopke:1988cs,Seiden:1988rr, Feldmann:1998vh, Haber:1985cv, Ball:1995zv, Bramon:1997mf, Gerard:1999uf, Zhao:2006gw, Li:2007ky, Escribano:2009jti,Morisita:1990cg}, where $P$ and $V$ represent light pseudoscalar and vector mesons, respectively. In these studies, the dominant part of the amplitude involves the annihilation of $c\bar{c}$ into light hadrons through the emission of three gluons, referred to as single-OZI (SOZI)  suppressed diagrams (Fig.~\ref{fd1-a}). Subsequently, double-OZI (DOZI) suppressed diagrams are introduced, as depicted in Figs.~\ref{fd1-b} and \ref{fd1-c}, where an additional gluon is exchanged between the final states. Furthermore, at leading order, the contribution from the electromagnetic (EM) diagrams  shown in Fig.~\ref{fd1-d} are limited~\cite{Seiden:1988rr,Li:2007ky}, which can be safely disregarded.
This approximation has successfully explained $J/\psi$ two-body decays. Along with the radiative decays of $J/\psi$ (Fig.~\ref{fd2}), this approximation effectively explains the mixing between $\eta$ and $\eta^{\prime}$ mesons and provides a way to investigate the mixings between $q\bar{q}$ and glueballs. In this work, we extend the study to $J/\psi \to S V$ processes to explore the mixing between isoscalar mesons and glueballs.

This paper is organized as follows. In Sec.~\ref{FORMALISM}, we present the
formalism of the decays in terms of $SU(3)_f$. In Sec.~\ref{Numerical Results}, we give our numerical results. We summarize this work in Sec.~\ref{Conclusion}.

\section{Formalism}\label{FORMALISM}

The 
$SU(3)_f$
representation of vector mesons
is given  as 
\begin{equation}
	\boldsymbol{V} = 
	\begin{pmatrix}
		\frac{\rho^0}{\sqrt{2}}+\frac{\omega}{\sqrt{2}}&\rho^+&K^{*+}\\
		\rho^-&\frac{-\rho^0}{\sqrt{2}}+\frac{\omega}{\sqrt{2}}&K^{*0}\\
		K^{*-}&\Bar{K}^{*0}&\phi\\
	\end{pmatrix}\,,
\end{equation}
where $\omega$ and $\phi$ are the physical states 
of $n\overline{n}$ and $s \overline{s}$, respectively.
Analogous to the vector mesons, the scalar meson nonet is defined  as 
\begin{equation}
	\small \boldsymbol{S}= 
	\begin{pmatrix}
		\frac{1}{\sqrt{2}}a_0^0+\frac{1}{\sqrt{6}}S_0^8 + \frac{1}{\sqrt{3}}S_0^1&a_0^+&K_0^{*+}\\
		a_0^-&-\frac{1}{\sqrt{2}}a_0^0+\frac{1}{\sqrt{6}}S_0^8+ \frac{1}{\sqrt{3}}S_0^1&K_0^{*0}\\
		K_0^{*-}&\Bar{K}_0^{*0}&-\frac{2}{\sqrt{6}}S_0^8+ \frac{1}{\sqrt{3}}S_0^1\\
	\end{pmatrix},
\end{equation}
where $S_0^8$ and  $S_0^1$ are octet and singlet states in the singlet-octet basis, respectively.
There is a widely used mix scheme, which includes $S_0^8$, $S_0^1$ and $G_0$, given by
\begin{equation}
	\label{eq2.3}
	\small
	\left(\begin{array}{ccc}
		f_0^1 \\
		f_0^2\\
		f_0^3
	\end{array}\right)_i
	=
	U_{ij}
	\left(\begin{array}{ccc}
		S_0^8 \\
		S_0^1 \\
		G_0
	\end{array}\right)_j = 
	\left( \begin{array}{ccc}
		c_{12}c_{13}   & s_{12}c_{13} & s_{13} \\
		-s_{12} c_{23} - c_{12} s_{23} s_{13} & c_{12} c_{23}- s_{12} s_{23} s_{13} & s_{23} c_{13}\\
		s_{12}s_{23} - c_{12}c_{23}s_{13} & - c_{12}s_{23} - s_{12}c_{23}s_{13} & c_{23} c_{13}
	\end{array} \right)_{ij}
	\left(\begin{array}{ccc}
		S_0^8 \\
		S_0^1 \\
		G_0
	\end{array}\right)_j  ,
\end{equation}
with $c_{ij} = \text{cos} \theta_{ij}$ and $s_{ij} = \text{sin} \theta_{ij}$. We choose $M_{f_0^3} > M_{f_0^2} > M_{f_0^1}$.

The amplitudes for the SOZI suppressed diagrams are expressed
in terms of the $SU(3)_f$ symmetry with the coupling strength $g$.
We assign the amplitude of the DOZI suppressed diagram in Fig.~\ref{fd1-b} (\ref{fd1-c}) restricted by
$r$ ($r^{\prime}$) of $O(10^{-1}$) comparing with the SOZI one.
Consequently, the effective Lagrangian density at leading order can be given by
\begin{equation}
\small
    \mathcal{L}^0_{SV} = \frac{g}{2}~\text{Tr}(\boldsymbol{S}\{\boldsymbol{V},\boldsymbol{S_g}\})+ rg\text{Tr}(\boldsymbol{S}\boldsymbol{S_g})\text{Tr}(\boldsymbol{V}\boldsymbol{S_g}) +r^{\prime}g\boldsymbol{G_0}\text{Tr}(\boldsymbol{V}\boldsymbol{S_g}),
\end{equation}
where $\boldsymbol{G_0}$ is a pure scalar glueball field, the matrices of $\boldsymbol{V}$ and $\boldsymbol{S}$ represent the vector and scalar mesons, respectively, and $\boldsymbol{S_g}
=$diag(1, 1, 1$-s$)$/\sqrt{3}$ is the
matrix of spurious fields corresponding to strong interactions.
Meanwhile, FIG.~\ref{fd2}  depicts the dominant processes of $J/\psi \to \gamma S$.  
The effective Lagrangian for $J/\psi$ radiative decays   is expressed as~\cite{Morisita:1990cg}
\begin{equation}
\small
    \mathcal{L}^0_{\gamma S} = d~\text{Tr}(\boldsymbol{S}\boldsymbol{S_g})+\frac{r^{\prime}}{r}d~\boldsymbol{G_0}+ \frac{f}{2}~\text{Tr}(\boldsymbol{S_g}\{\boldsymbol{S},\boldsymbol{S_e}\}),
\end{equation}
where $\boldsymbol{S_e}
= $diag($2,-1,-1$)/3 is spurious fields corresponding to electromagnetic interaction, and the first, second and third terms come from  FIGs.~\ref{fd2-a}, \ref{fd2-b} and \ref{fd2-c}, respectively.

Finally,
the decay widths of $J/\psi \to SV $ and $\gamma S$ are given by~\cite{Morisita:1990cg}
\begin{equation}
\small
    \Gamma(J/\psi \to SV, \gamma S)  = \frac{p^3}{12\pi m_{J/\psi}^2} |\mathcal{A}|^2,
\end{equation}
where $p$ is the center-of-mass momentum of $S$. The amplitudes for the decays $J/\psi~\to~VS\ \text{and}\ \gamma S$ are tabulated in Table~\ref{amplitudes for J/Psi}.

\begin{table}[!htb]
  \small
  \caption{Amplitudes for the decays of $J/\psi \to S V$ and $\gamma S$.}
  \label{amplitudes for J/Psi}
  \centering
  \begin{tabular}{c|c}
  \hline
  \hline
  Channels & Amplitude \\
  \hline
  $K^*(S)K^*(V)$ &   $\frac{1}{\sqrt{3}}g$  
  \\
  $\rho a_0$ &        $\frac{1}{\sqrt{3}}(1-\frac{1}{2}s)g$ 
  \\
    $\omega f_0^i$ &   $ \left(\frac{1}{\sqrt{3}}g + \frac{2}{3\sqrt{3}}srg \right) \cdot U_{i1} +
    \left(\frac{\sqrt{2}}{3}g+\frac{\sqrt{2}}{\sqrt{3}}(1-s)rg \right) \cdot U_{i2} + \frac{\sqrt{2}}{\sqrt{3}}r^{\prime}g \cdot U_{i3}$  
  \\
  $\phi f_0^i$  &    $\left(-\frac{\sqrt{2}}{3}g+\frac{\sqrt{2}}{3\sqrt{3}}srg\right)(1-s)\cdot U_{i1} +
  \left(\frac{1}{3}g + \frac{1}{\sqrt{3}}(1-\frac{1}{3}s)rg\right)(1-s)\cdot U_{i2} + 
  \frac{1}{\sqrt{3}}r^{\prime}g(1-s)\cdot U_{i3}$
  \\

    $\gamma f_0^i$ & $\left(\frac{2}{\sqrt{3}}sd+\frac{\sqrt{2}}{6}(1-\frac{1}{3}s)f\right)\cdot U_{i1}
    + \left((1-\frac{1}{3}s)d+\frac{1}{9}sf\right)\cdot U_{i2} + \frac{r^{\prime}}{r}d\cdot U_{i3}$
  \\

    $\gamma a_0$ & $\frac{1}{\sqrt{6}}f$
  \\
  \hline
  \hline
  \end{tabular}
\end{table}

\section{Numerical Results}\label{Numerical Results}
The scalar meson spectroscopy, presented in Table \ref{groups based on a0}, aids in identifying and classifying mesons into distinct nonets based on the masses of the $a_0$ mesons. We determine the coupling strengths and the mixing angles by using the experimental data for the $J/\psi \to SV$ and $\gamma S$ decays.
For $J/\psi \to SV$, we rely on the data collected from BESII \cite{BES:2004zql, Wu:2001vz, BES:2004wqz, BES:2004twe}. Regarding $J/\psi \to \gamma S$, we use those compiled in Ref.~\cite{Sarantsev:2021ein}.

\begin{table}[!htb]
    \caption{Scalar meson menu. The mixing of $q\bar{q}$ with glueball in $S(1450)$ is under consideration. The existence of $S(1710)$ is postulated as a natural consequence of the presence of $a_0(1710)$ and $K^*(1950)$. 
 }
    \label{groups based on a0}
    \centering
    \renewcommand\arraystretch{1}
    \tabcolsep=0.35cm
    \begin{tabular}{c|ccccc}
        \hline
        \hline
           &   $S(980)$   &    $S(1450)$  &   $S(1710)$   \\ \hline
           & $a_0(980)$ & $a_0(1450)$ & $a_0(1710)$ 
           \\ 
           & $K^*(700)$ & $K^*(1430)$ &   $K^*(1950)$     
\\ 
\hline
         $f_0^1$    & $f_0(500)$ & $f_0(1370)$ & \footnote{\raggedright 
         The recurrence of $f(1710)$ in $S(1450)$ and $S(1710)$ is attributed to testing hypothesis.
         }$f_0(1710)$  \\
         $f_0^2$        & $f_0(980)$ & $f_0(1500)$ & $f_0(1770)$  \\ 
         $f_0^3$        &             &   $f_0(1710)$             \\  
         \hline \hline
    \end{tabular}
\end{table}

\subsection{The nonet below 1 GeV{}}
We start with the low mass scalar mesons, focusing on the nonet denoted as $S(980)$. To explain the experiment, the mixings between $q\bar{q}$ and glueballs are not considered.   Our results, along with the experimental data, are shown in Table~\ref{S980 data}.

The experimental upper limit of  $B(J/\psi \to \gamma a_0 \to \gamma \eta \pi^0) < 2.5\times10^{-6}$~\cite{BESIII:2016gkg}, indicating that $|f| < 0.52\times10^{-3}$. In this range of values, the impact of $f$ on other decay channels is minor.
Consequently, this limitation is added to the parameter $f$. The obtained results are given by
\begin{equation}
\begin{aligned}
    &g=(9.22\pm 1.20) \times10^{-3}, ~~~~r = -(0.166\pm0.083), ~~~~ s = 0.02\pm0.15,\\ 
    &\theta_{12}  = (82.9\pm4.4)^{\circ}, ~~~~~~~~d = (3.41\pm0.37)\times10^{-3}, ~~~~f = -0.52\times10^{-3},
\end{aligned}
\end{equation}
with $\chi^2 = 0.72$ for 2 degrees of freedom.
	This implies that $f_0(500)$ exhibits a strong tendency towards being a singlet state, while $f_0(980)$ is closer to an octet state, which is also supported by the BESII data~\cite{BES:2004twe} of $f_0 \to \pi \pi , K \overline{K}$~\cite{Anisovich:2011zz}.

\begin{table}[!htb]
  \caption{The experimental and fitting branch ratios of $J/\psi \to S(980) V$ in units of $10^{-4}$.}
  \label{S980 data}
  \centering
  \renewcommand\arraystretch{1}
  \tabcolsep=0.35cm
  \begin{tabular}{cccccc}
  \hline
  \hline
  &Channels    &Data  & This work   \\
  \hline

  &$\omega f_0(980)$  &   $5.4\pm1.8$~\cite{Wu:2001vz}   & $ 4.3\pm0.9$   
  \\
  \hline
  &$\phi f_0(980)$    &  $9.9\pm1.7$  &    $10.0\pm1.5$  
  \\
  \hline
   &$\omega f_0(500)$    &$11.7\pm7.3$~\cite{BES:2004wqz}      &      $10.0\pm7.0$   
   \\
  \hline
  &$\phi f_0(500)$   &$1.8\pm0.7$   &$1.8\pm0.7$ \\
  \hline
  &$K^*(892)^{\pm}K^*(700)^{\mp}$      &$11^{+10}_{-6}$  &   $17.5\pm2.8 $ 
  \\
  \hline
  &$\rho a_0(980)$    &$-$    & $17.5\pm4.6 $ 
 
  \\
  \hline
  &$\gamma f_0(980)$ & $0.21\pm0.04$ &  $0.21\pm0.04 $ \\
  \hline
  &$\gamma f_0(500)$ &$11.4\pm2.1$  &   $11.5\pm2.1$ \\

  \hline
  \hline
    \end{tabular}
\end{table}

\begin{table}[!htb]
  \caption{ The mixing angle of $\theta_{12}$ in $S(980)$ along with the ones in the literature. 
 }
  \label{mixing angle}
  \centering
  \begin{threeparttable}
   \begin{tabular}{c|c|c}
  \hline
  \hline
  This work &   $q\bar{q}$ model &  Tetraquark model  \\ \hline 
  $(82.9\pm4.4)^{\circ}$ &  \tnote{1}\ $(71\pm5)^{\circ}$~\cite{Oller:2003vf}, \tnote{1}\ $(82\pm1.2)^{\circ}$~\cite{Klempt:2021nuf}, $54.7\sim 71.7^{\circ}$~\cite{LHCb:2014vbo}, &$59.7^{\circ}$~\cite{tHooft:2008rus},
  \\
   &$(88.7^{+9}_{-15})^{\circ}$\cite{Li:2012sw}, $(77.7\pm4)^{\circ}$~\cite{LHCb:2015klp}&$(66.7\pm3.5)^{\circ}$~\cite{LHCb:2015klp}
  \\
  \hline
  \hline
  \end{tabular}
  \begin{tablenotes}
      \footnotesize
      \item[1]{The mixing angle reported is corresponded to $\theta = 90^{\circ}-\theta^{\prime}$.}
   \end{tablenotes}
  \end{threeparttable}
\end{table}

We present our results of the mixing angle along with those predicted in the literature in Table~\ref{mixing angle}. 
Ref.~\cite{Oller:2003vf} is based on the $q\bar{q}$ model and an $SU(3)_f$ analysis of meson decays, yielding $\theta = (71\pm5)^{\circ}$. 
In Ref.~\cite{Klempt:2021nuf}, the radiative decays of $J/\psi \to \gamma f_0(980,\ 500)$ were used and found $\theta = \pm(82\pm1.2)^{\circ}$.
The LHCb collaboration~\cite{LHCb:2014vbo} investigated $\bar{B^0} \to J/\psi \pi^+ \pi^-$.
Ref.~\cite{Li:2012sw} studied $B^0_s \to J/\psi f_0(980)$ with $\theta = (89^{+9}_{-15})^{\circ}$.
Through Dalitz plot analyses of $B^0 \to \bar{D}^0\pi^+\pi^-$, Ref.~\cite{LHCb:2015klp} obtained $\theta = (77.7\pm4)^{\circ}$. 

In the tetraquark model, by studying $S \to PP$,  Ref.~\cite{tHooft:2008rus} argued that the low-lying scalar mesons are tetraquark states and determined the mixing angle to be $54.7^{\circ}<\theta<59.7^{\circ}$. Ref.~\cite{LHCb:2015klp} also provided a solution within the tetraquark model, yielding $\theta = (66.7\pm3.5)^{\circ}$.

The tetraquark hypothesis exhibits smaller mixing angles that are insufficient to explain the experimental data.
Besides, we emphasise the need for further experimental studies to reduce uncertainties of channels $J/\psi \to \omega f_0(500)$ and $K^*(892)^{\pm}K^*(700)^{\mp}$.

\subsection{The nonets in range of  1-2~GeV}\label{3B}
In this section, $f_0(1370,\ 1500,\ 1710)$ are considered as members of $S(1450)$ and 
mixtures of $S_0^8$, $S_0^1$, and $G_0$ as shown in Eq.~(\ref{eq2.3}). 
Given that the contribution from the three-gluon-annihilation diagram is small, 
whether in $\gamma P$ final states~\cite{Morisita:1990cg} or the $\gamma S$ processes 
mentioned above, we assume that $f$ possesses the same value as in $S(980)$. 
Additionally, the breaking terms fluctuate within the range of $-0.3$ to $0.3$, 
exerting an impact on the subsequent numerical results of less than $10\%$. 
Since the breaking term $s \sim 0$ in $S(980)$, we set it to 0 here. 
Consequently, we can use 7 available data points to determine 7 parameters, obtained as 
\begin{equation}
  \begin{aligned}
      &~~~~~~~~~~~~~~g=(1.14\pm 0.18) \times10^{-2}, ~~~~~~~~~~r = -(0.716\pm0.291), \\ 
      &~~~~~~~~~~~~~~d = (4.52\pm1.07)\times10^{-3}, ~~~~~~~~~~r^{\prime} = -(0.633\pm0.132),\\
      &~~~~~~~\theta_{12} = (51.4\pm6.4)^{\circ},~~~~~~~\theta_{13} = (-0.1\pm14.6)^{\circ},~~~~~~~\theta_{23} = (2.3\pm8.6)^{\circ}.
  \end{aligned}
  \end{equation}
The comparisons between our fitting results and the experimental data are presented 
in Table~\ref{CKM scheme}.
It is important to highlight that the parameter $f$ can be determined by the decay
$J/\psi \to \gamma a_0(1450)$, which  varies only  on  $f$.
Furthermore, we notice that $r^{\prime}/r \sim 1$, 
which implies that the contributions in Figs.~\ref{fd1-b} and 
~\ref{fd1-c}, as well as Figs.~\ref{fd2-b} and~\ref{fd2-c}, are of the same order,
consistent with Refs.~\cite{Morisita:1990cg,Li:2007ky}.
\begin{table}[!htb]
  \caption{The experimental and fitting branch ratios  of $J/\psi \to S(1450) V, S(1450) V$ in units of $10^{-4}$. }
  \label{CKM scheme}
  \centering
  \renewcommand\arraystretch{1}
  \tabcolsep=0.35cm
  \begin{tabular}{cccc}
  \hline
  \hline
  &Channels     &Data   & this work   \\
  \hline

   &$\omega f_0(1370)$    &$-$      &      $0.7^{+1.3}_{-0.7}$   \\
  \hline
   &$\omega f_0(1500)$  &   $-$   &  $5.1\pm3.5$ \\
   \hline
   &$\omega f_0(1710)$  &   $6.6\pm1.3$ ~\cite{BES:2004zql}   &  $6.6\pm1.3$\\
   \hline
  &$\phi f_0(1370)$   &$4.6\pm1.4$  &   $4.6\pm1.4$ \\
  \hline
  &$\phi f_0(1500)$    &  $2.5\pm1.3$  &   $2.5\pm1.3$     \\
  \hline
  &$\phi f_0(1710)$    &  $2.0\pm 0.7$  &   $2.0\pm 0.7$    \\
  \hline
  &$\gamma f_0(1370)$ &$6.9\pm1.2$  &   $6.9\pm1.2$  \\
  \hline
  &$\gamma f_0(1500)$ & $4.7\pm0.9$ &  $4.7\pm0.9$ \\
  \hline
  &$\gamma f_0(1710)$ & $5.6\pm1.0$ &  $5.6\pm1.0$ \\
  \hline
  &$K^*(892)^{\pm}K^*(1430)^{\mp}$      &$-$  &   $13.2\pm4.3$ \\
  \hline
  &$\rho a_0(1450)$    &$-$    & $15.0\pm4.9$ \\
  \hline
  &$\gamma a_0(1450)$ & $-$ &  $0.024$ \\
  \hline
  \hline
  \end{tabular}

\end{table}

The mixing angle $\tan  \theta_{12} \approx  \sqrt{2} $ implies that 
 $f_0(1370)$ and $f_0(1500)$ in $S(1450)$ 
play the  roles of $\omega$ and $\phi$ in $\boldsymbol{V}$. Meanwhile, $\theta_{13,23} \approx 0$ suggest the nature of $f_0(1710)$ being a glueball.
This interpretation aligns with the findings from $J/\psi$ radiative decays~\cite{Sarantsev:2021ein,Gui:2012gx}
and the BESIII data~\cite{BESIII:2022riz,BESIII:2022iwi,Brunner:2015oga}.


Recently, the BaBar collaboration reported the observation of the scalar meson $a_0(1710)$~\cite{BaBar:2021fkz}, subsequently confirmed by the BESIII collaboration~\cite{BESIII:2022npc}. 
Alongside the discovery of $f_0(1770)$ and $K^*(1950)$, an additional $SU(3)_f$ nonet is expected. 
For instance, we consider the grouping of 
$
S(1710) \ni
\{
f_0(1710), f_0(1770) , a_0(1710), K^*(1950)
\} $. 
Under this grouping scenario, $B(J/\psi \to \gamma f_0(1710), \gamma f_0 ( 1770) ) = (5.6\pm1.0, 18.1 \pm 2.6 ) \times 10^{-4}$
implies 
$f_0(1710)$ and $f_0(1770)$ belong to $S_0^8$ and $S_0^1$, respectively, which conflicts with the BESII data~\cite{BES:2004twe,Ablikim:2006db,Klempt:2021wpg}.
The discrepancy shows that $f_0(1710)$ does not belong to $S(1710)$ and is a plausible glueball candidate. 
However, due to limited experimental data, 
a comprehensive analysis containing the mixture of glueball and $S(1710)$ is not yet available. 
 
It is interesting to point out that the presence of $a_0(1950)$, $K^*(2130)$, and several $f_0$ particles implies the existence of an $SU(3)_f$ nonet around 2 GeV. In principle, the method studied in this work can also apply to them. Nevertheless, a numerical analysis cannot be carried out at the current stage due to the limited data points. To further understand the nature of these exotic states, we urge our experimental colleagues to measure the decays of $J/\psi$ with $f_0$ in the final states.

\section{Conclusion}\label{Conclusion}
We  have examined  the    nonet of scalar mesons below 1 GeV using the data of  $J/\psi \to {S}{V}, \gamma {S} $. The singlet-octet mixing angle 
between $f_0(500)$ and $f_0(980)$ has been determined as 
$\theta_{12} = (83.5\pm 1.0)^{\circ}$, which is consistent with the calculations  within  the $q\bar{q}$ hypothesis.

For the scalar mesons in the range of 1-2 GeV, we have considered  $f_0(1370,\ 1500,\ 1710)$ are  the mixture of  $q\bar{q}$ and $G_0$. 
With $a_0(1450)$ and  $K^*(1430)$, together they form an $SU(3)_f$ nonet plus a glueball. 
According to the data of $J/\psi$ decays, we have found  that $f_0(1370)$, $f_0(1500)$ and $f_0(1710)$ are mainly $n\bar{n}$, $s\bar{s}$ and $G_0$, respectively. 
In the same mass range, 
we have studied the assumption of 
 an additional  nonet, containing
$a_0(1710)$, 
$f_0(1710)$ and $f_0(1770)$ as its components. We have found that this assumption is not compatible with 
the  current data.
It reinforces the conclusion of  $f_0(1710)$ containing a large glueball component.

We recommend the future experiments to measure $J/\psi \to \rho a_0(980,\ 1450,\ 1710), $ $ K^*(892)^{\pm} K^*(700,\ 1430,\ 1950)^{\mp}
$ and $\omega f_0(500)$, as our $SU(3)_f$ fit is sensitive to them. Through these measurements, more accurate determinations of the mixing angles and the constituents of different nonets can be obtained.

\acknowledgments
We would like to express our sincere appreciation to Jiabao Zhang for his valuable insights during the development of this work. This work is supported in part by the National Key Research and Development Program of China under Grant No. 2020YFC2201501 and  the National Natural Science Foundation of China (NSFC) under Grant No. 12347103 and 12205063.

  \end{document}